# An Intelligent Network Selection Strategy Based on MADM Methods in Heterogeneous Networks

Lahby Mohamed[1], Cherkaoui Leghris[2] and Adib Abdellah[3]

[1,2,3]Computer Science Department, LIM Lab
Faculty of Sciences and Technology of Mohammedia,
B.P. 146 Mohammedia, Morocco
Email: {mlahby, cleghris, adib_adbe}@yahoo.fr

## ABSTRACT

*Providing service continuity to the end users with best quality is a very important issue in the next generation wireless communications. With the evolution of the mobile devices towards a multimode architecture and the coexistence of multitude of radio access technologies (RAT's), the users are able to benefit simultaneously from these RAT's. However, the major issue in heterogeneous wireless communications is how to choose the most suitable access network for mobile's user which can be used as long as possible for communication.*

*To achieve this issue, this paper proposes an intelligent network selection strategy which combines two multi attribute decision making (MADM) methods such as analytic network process (ANP) and the technique for order preference by similarity to an ideal solution (TOPSIS) method. The ANP method is used to find the differentiate weights of available networks by considering each criterion and the TOPSIS method is applied to rank the alternatives. Our new strategy for network selection can dealing with the limitations of MADM methods which are the ranking abnormality and the ping-ponf effect.*

## KEYWORDS

*Heterogeneous Wireless Network, Network Selection, Multi Attribute Decision Making, Ranking Abnormality, Ping Pong Effect.*

## 1. INTRODUCTION

In recent years, the next generation wireless communications are growing rapidly and are integrating a multitude of radio access technologies (RAT's) such as wireless technologies (802.11a, 802.11b, 802.15, 802.16, etc.) and cellular networks (GPRS, UMTS, HSDPA, LTE, etc.). With the evolution of the mobile devices towards a multimode architecture and the coexistence of these heterogeneous RAT's the users are able to benefit simultaneously from these RAT's and they can also use various services offered by each type of access network.

However the most important issue in RAT's, is to provide ubiquitous access for the end users, under the principle "Always Best Connected" (ABC) [1], to achieve this issue a vertical handoff decision [2] is intended to determine whether a vertical handoff should be initiated, and to choose the most suitable network in terms of quality of service (QoS) for mobile users. The handover vertical process can be divided into three steps:

1) Handover initiation: it contains some preparation for handoff such as the measurement of received signal strength (RSS), QoS, security, battery level, etc.
2) Handover decision: it consists on choosing the most suitable network access among those available to perform a handover.
3) Handover execution: it consists on establishing the target access network by using mobile IP protocol (MIP).

The network selection problem is the most important key of the handover vertical decision. For that, our work focuses on the optimization of the network selection decision for users in order to support many services with best QoS and let the users stay connected with the current access network as long as possible. However, no single wireless network technology is considered to be more favorable than other technologies in terms of QoS. In other words, each network access in RAT's seems to be specifically characterized by the bandwidth offered, the coverage ensured by the network as well as the cost to deliver the service. Moreover, there is some kind of complementarity between these various networks, for example, 801.11a offers a higher bandwidth with a cover limited, while UMTS ensures a large cover with lower bandwidth. The network selection algorithm depends on multiple criteria which are:

- From terminal side: battery, velocity, etc.
- From service side: QoS level, security level, etc.
- From network side: provider's profile, current QoS parameters, etc.
- From user side: users preferences, perceived QoS, etc.

In the other hand the network selection problem can be tackled with several schemes and decision algorithms such as genetic algorithms [3], fuzzy logic [4], utility functions [5] and multi attribute decision making (MADM) methods [6,7,8,9,10,11,12]. In [3] the genetic algorithm is applied to optimize the access network function with the goal of selecting the optimal access network. In [4] the authors have proposed an intelligent approach for vertical handover based on fuzzy logic. In [5] the authors proposed a network selection scheme based on utility function which takes more key factors for multimedia communication in the future urban road wireless networks. These factors include data rate, bit error rate, latency, power consumption, monetary cost, load balance, individual's preference and handoff stability.

Due to great number of criteria and algorithms which can be used in network selection, the most challenging problems focus in selecting the appropriate criteria and definition of a strategy which can exploit these criteria. According to nature of network selection problem, MADM algorithms represent a promising solution to select the most suitable network in terms of quality of service (QoS) for mobile users. However the major limitations of MADM methods are the ranking abnormality and the ping-pong effect. The ranking abnormality means that the ranking of candidate networks change when low ranking alternatives are removed from the candidate list, which can make the selection problem inefficient. The ping pong effect occurs when the terminal mobile performs excessive handoffs for a given time which causing the higher number of handoffs. This phenomenon can led to increasing in power consumption and the decreasing in throughput.

To address the limitations posed by MADM methods, we propose an intelligent network selection strategy based on analytical network process (ANP) and the technique for order preference by similarity to an ideal solution (TOPSIS) method, the ANP method is applied to find the weights of each criterion and TOPSIS method is used to rank the alternatives. The intelligence of our strategy focuses in two aspects: firstly we utilize the differentiate weights of available networks by considering each criterion in order to reduce the ranking abnormality and secondly we introduce the history criterion to reduce the number of handoff and to ensure that the terminal mobile stay connected to the current access network as long as possible.

This paper is organized as follows. Section 2 presents review of related work concerning network selection decision based on MADM methods. Section 3 describes multi attribute decision making methods (MADM). Section 4 presents our access network selection algorithm based on ANP and TOPSIS two MADM methods. Section 5 includes the simulations and results. Section 6 concludes this paper.

## 2. RELATED WORK

The MADM methods represent promising solution for solving the network selection problem. The MADM includes many methods such as analytic hierarchy process (AHP), analytic network process (ANP), simple additive weighting (SAW), multiplicative exponential weighting (MEW), grey relational analysis (GRA), technique for order preference by similarity to ideal solution (TOPSIS) and the distance to the ideal alternative (DIA). In [6] comparison of network selection algorithms, between two methods which are the hybrid ANP algorithm and Blume algorithm is proposed. The hybrid ANP approach combines two MADM methods such as ANP method and rank reversal TOPSIS (RTOPSIS). The ANP method is used to get weights of the criteria and RTOPSIS method is applied to determine the ranking of access network. In [7] and [8], the network selection algorithm is based on AHP and GRA, the AHP method is used to determine weights for each criterion and GRA method is applied to rank the alternatives. In [9], [10] and [11] the network selection algorithm combines the AHP method and the TOPSIS method, the AHP method is used to get weights of the criteria and TOPSIS method is applied to determine the ranking of access network.

Among MADM methods mentioned above, TOPSIS method has been extensively used to solve the network selection problem. However, TOPSIS still suffers from ranking abnormality, some proposals were presented to avoid this issue, in [9] the author has proposed an iterative approach for application of TOPSIS for network selection problem. The disadvantage of this method lies in the computation time, for example, if we have n available access networks we must repeat iterative TOPSIS n-1 until the best interface network is reached. Reference [12] presents DIA algorithm which selects the alternative that is the shortest euclidean distance to positive ideal alternative. One of the main disadvantages of DIA method is doesn't take into account the normalization type, in other words, when the low ranking alternative is removed from the candidate list, the normalized attribute values of all alternatives will be changed and the ranking order of the alternative will be changed as well. Another disadvantage of this method is that, the euclidean distance used by DIA doesn't take into consideration the correlation between different criteria, all the components of the vectors will be treated in the same way.

The major factor causing the ranking abnormality is the weighting algorithm [13] used to weigh different criteria, in addition the all decision algorithms based on MADM methods use the same weight vector of the all available networks, in the other words each algorithm for network selection decision don't take into account the user preference relative to each access network according to each criterion. Due to the criteria are the same relative importance in each access network in the classical network selection algorithms, in our new strategy the ANP method is applied to find the differentiate weights of available networks by considering each criterion.

On the other hand the all selection decision algorithms based on MADM methods mentioned above still suffer from the ping-pong effect, to cope with this issue we introduce the history criterion to reduce the number of handoff and to ensure that the terminal mobile stay connected to the current access network as long as possible.

## 3. MULTI- ATTRIBUTE DECISION MAKING

### 3.1. ANP

The analytic network process (ANP) is a MADM method, proposed by Saaty [14], which extends the AHP approach to problems with dependence and feed beck within clusters (inner dependence) and between clusters (outer dependence). The ANP approach is based on six steps:

1) Model construction: A problem is decomposed into a network in which nodes corresponds to components. The elements in a component can interact with some or all of the elements of another component. Also, relationships among elements in the same component can exist. These relationships are represented by arcs with directions.

2) Construct of the pairwise comparisons: To establish a decision, ANP builds the pairwise matrix comparison such as:

$$A = \begin{bmatrix} x_{11} & x_{12} & \cdots & \cdots & x_{1n} \\ x_{21} & x_{22} & \cdots & \cdots & x_{2n} \\ \vdots & \vdots & \ddots & \vdots & \vdots \\ \vdots & \vdots & \vdots & \ddots & \vdots \\ x_{n1} & x_{n2} & \cdots & \cdots & x_{nn} \end{bmatrix}, where \begin{cases} x_{ii} = 1 \\ x_{ji} = \dfrac{1}{x_{ij}} \end{cases} \quad (1)$$

Elements $x_{ij}$ are obtained from the table 1, it contains the preference scales.

Table 1: Saaty's scale for pair-wise comparison

| Saaty's scale | The relative importance of the two sub-elements |
|---|---|
| 1 | Equally important |
| 3 | Moderately important with one over another |
| 5 | Strongly important |
| 7 | Very Strongly important |
| 9 | Extermely important |
| 2,4,6,8 | Intermediate values |

3) Construct the normalized decision matrix: $A_{norm}$ is the normalized matrix of A(1), where $A(x_{ij})$ is given by, $A_{norm}(a_{ij})$ such:

$$A_{norm} = \begin{bmatrix} r_{11} & r_{12} & \cdots & \cdots & r_{1n} \\ r_{21} & r_{22} & \cdots & \cdots & r_{2n} \\ \vdots & \vdots & \ddots & \vdots & \vdots \\ \vdots & \vdots & \vdots & \ddots & \vdots \\ r_{n1} & r_{n2} & \cdots & \cdots & r_{nn} \end{bmatrix}, where \; r_{ij} = \frac{x_{ij}}{\sum_{i=1}^{n} x_{ij}} \quad (2)$$

4) Calculating the weights of criterion: The weights of the decision factor i can be calculated by:

$$W_i = \frac{\sum_{j=1}^{n} a_{ij}}{n} \; and \; \sum_{j=1}^{n} W_i = 1 \quad (3)$$

With n is the number of the compared elements.

5) Calculating the coherence ratio (CR): To test consistency of a pairwise comparison, a consistency ratio (CR) can be introduced with consistency index (CI) and random index (RI).

- Let define consistency index CI

$$CI = \frac{\lambda_{max} - n}{n - 1} \quad (4)$$

- Also, we need to calculate the $\lambda_{max}$ by the following formula:

International Journal of Wireless & Mobile Networks (IJWMN) Vol. 4, No. 1, February 2012

$$\lambda_{max} = \frac{\sum_{j=1}^{n} b_i}{n}, \quad where \quad b_i = \frac{\sum_{j=1}^{n} W_i * a_{ij}}{W_i} \quad (5)$$

- We calculate the coherence ratio CR by the following formula

$$CR = \frac{CI}{RI} \quad (6)$$

The various values of RI are shown in table 2.

Table 2: value of random consistency index RI

| Criteria | 3 | 4 | 5 | 6 | 7 | 8 | 9 | 10 |
|---|---|---|---|---|---|---|---|---|
| RI | 0.58 | 0.90 | 1.12 | 1.24 | 1.32 | 1.41 | 1.45 | 1.49 |

If the CR is less than 0.1, the pairwise comparison is considered acceptable.

6) Construct the super-matrix formation: The local priority vectors are entered into the appropriate columns of a super-matrix, which is a partitioned matrix where each segment represents a relationship between two components.

## 3.2. TOPSIS

Technique for order preferences by similarity to an ideal solution (TOPSIS), known as a classical multiple attribute decision-making (MADM) method, has been developed in 1981 [15]. In TOPSIS method, the optimal alternative selected should have the shortest distance from the positive ideal solution and the farthest distance from the negative ideal solution. The procedure can be categorized in six steps:

1) Construct of the decision matrix: the decision matrix is expressed as

$$D = \begin{bmatrix} d_{11} & d_{12} & \cdots & \cdots & d_{1m} \\ d_{21} & d_{22} & \cdots & \cdots & d_{2m} \\ \vdots & \vdots & \ddots & \vdots & \vdots \\ \vdots & \vdots & \vdots & \ddots & \vdots \\ d_{n1} & d_{n2} & \cdots & \cdots & d_{nm} \end{bmatrix} \quad (7)$$

Where $d_{ij}$ is the rating of the alternative Ai with respect to the criterion $C_j$

2) Construct the normalized decision matrix: each element $r_{ij}$ is obtained by the Euclidean normalization;

$$r_{ij} = \frac{d_{ij}}{\sqrt{\sum_{i=1}^{m} d_{ij}^2}}, \text{ i=1,…,m and j=1,…,n.} \quad (8)$$

3) Construct the weighted normalized decision matrix: The weighted normalized decision matrix $v_{ij}$ is computed as:

$$v_{ij} = W_i * r_{ij} \quad where \quad \sum_{j=1}^{n} W_i = 1 \quad (9)$$

4) Determination of the ideal solution $A^*$ and the anti-ideal solution $A^-$:

$$A^* = [V_1^*, …, V_m^*] \quad and \quad A^- = [V_1^-, …, V_m^-] \quad (10)$$

- For desirable criteria:

$$V_i^* = \max\{v_{ij}, j = 1, \ldots, n\} \quad (11)$$

$$V_i^- = \min\{v_{ij}, j = 1, \ldots, n\} \quad (12)$$

- For undesirable criteria:

$$V_i^* = \min\{v_{ij}, j = 1, \ldots, n\} \quad (13)$$

$$V_i^- = \max\{v_{ij}, j = 1, \ldots, n\} \quad (14)$$

5) Calculation of the similarity distance:

$$S_j^* = \sqrt{\sum_{j=1}^{m}(V_i^* - v_{ij})^2}, j = 1, \ldots, n \quad (15)$$

And

$$S_j^- = \sqrt{\sum_{j=1}^{m}(V_i^- - v_{ij})^2}, j = 1, \ldots, n \quad (16)$$

6) Ranking:

$$C_j^* = \frac{S_j^-}{S_j^* + S_j^-}, j = 1, \ldots, n \quad (17)$$

A set of alternatives can be ranked according to the decreasing order of $C_j^*$.

## 4. ACCESS NETWORK SELECTION STRATEGY

In order to deal with the ranking abnormality and to reduce the number of handoffs, we propose new intelligent network selection strategy based on two MADM methods such as ANP method and TOPSIS method. The ANP method is applied to find the weights of available networks by considering each criterion and TOPSIS method is applied to determine the ranking of each access network. Moreover our strategy introduces a new criterion namely history. This attribute allows to memorise the overall score given to the available network by using the TOPSIS method (history value is $C_j^*$).

The algorithm assumes wireless overlay networks which entail three heterogeneous networks such as UMTS, WLAN and WIMAX. Instead of using six attributes associated in this heterogeneous environment which are: Cost per Byte (CB), Available Bandwidth (AB), Security (S), Packet Delay (D), Packet Jitter (J) and Packet Loss (L), we add a new history criterion (H). Due to relationships between the QoS parameters such as AB, D, J and L, and based on survey and comparison study on weighting algorithms for access network selection presented in [13], the ANP method is the most appropriate algorithm which can be used to assign weights for each criterion.

Figure 1. exhibits the three levels based on ANP hierarchy for our new network selection strategy which takes into consideration the history attribute. The level 1 includes four criteria QoS, security, cost and history, the level 2 includes four QoS parameters such as AB, D, J and L and the level 3 includes three available networks UTMS, WIFI and WIMAX.

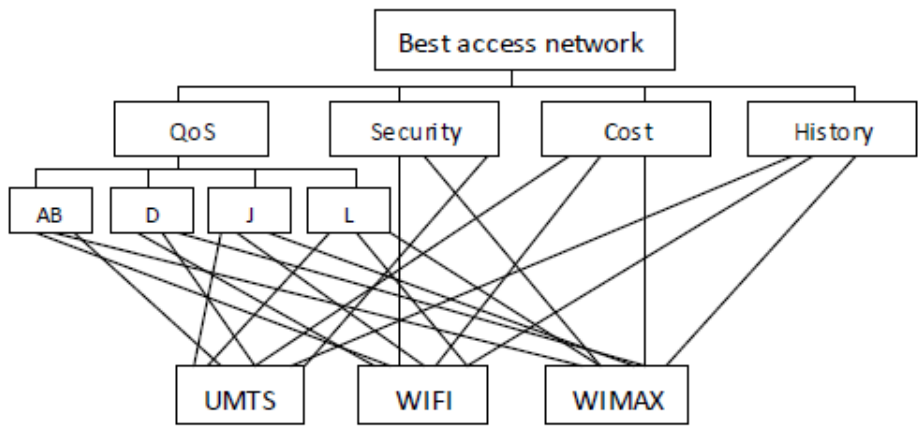

Figure 1. ANP hierarchy for our network selection problem

Based on the specific characteristics of the traffic type [16], our new strategy can be categorized in five steps:

1) Assign weights to level-1-criteria: the ANP method is used to get a weight of the decision criteria of level 1.
2) Assign weights to level-2-criteria: the ANP method is used to get a weight of the decision criteria of level 2 and to eliminate the interdependence impact of QoS sub-criteria.
3) Assign weights to level-3-alternatives: the ANP method is used to find the weights of the available networks by considering each criterion.
4) Obtain the vector weights of each available network: each access network will have dissimilar unique weights vector which will differ from those of other available networks, the weight vector of each available network is calculated by multiplication of the weight vector obtained in level 1 with the weight vector obtained in level 2 and with the weight vector obtained in level 3.
5) Select the best access network: the method TOPSIS is applied to rank the available networks and select the access network that has the highest value of $C_j^*$ (see the steps of TOPSIS method).

## 5. SIMULATION AND RESULTS

In order to illustrate the effectiveness of our new strategy based on ANP and TOPSIS which taking into consideration the user preference relative to each access network according to each criterion and including a new history attribute, we present performance comparison between four algorithms namely:

- TOPSIS-1: the network selection algorithm combines ANP method and TOPSIS method without considering differentiated weight of criterion and without considering the history attribute.
- TOPSIS-2: the network selection approach is based on ANP and TOPSIS and taking on consideration only the history attribute.
- TOPSIS-3: the network selection algorithm is based on ANP and TOPSIS and taking on consideration only the differentiated weight of criterion according to specific access network.

- TOPSIS-4: it's our new network selection strategy based on ANP and TOPSIS which considering differentiated weight of criterion according to specific access network and including the history attribute.

We simulate four traffic classes [16] namely background, conversational, interactive and streaming. In each simulation the four algorithms were run in 12 vertical handoff decision points and the performance evaluation is focused on two aspects, which are ranking abnormality and number of handoffs. For TOPSIS-1 and TOPSIS-3 the history criterion for each access network has no effect on our simulation.

Table 3: Attribute values for the candidate networks

| Criteria Network | CB (%) | S (%) | AB (mbps) | D (ms) | J (ms) | L (per$10^6$) | H (%) |
| --- | --- | --- | --- | --- | --- | --- | --- |
| UMTS | 60 | 70 | 0.1-2 | 25-50 | 5-10 | 20-80 | 100 |
| WLAN | 10 | 50 | 1-11 | 100-150 | 10-20 | 20-80 | 100 |
| WIMAX | 40 | 60 | 1-60 | 60-100 | 3-10 | 20-80 | 100 |

During the simulation, for each candidate networks, the measures of six attributes CB, AB, S, D, J and L are randomly varied according to the ranges shown in table 3. Furthermore the value of history criterion is initialized by 1, after the value of $H_{i+1}$ is equal to $C_j^*$ in iteration i+1 where $C_j^*$ is the score of TOPSIS method obtained in iteration i.

## 5.1. Simulation 1

In this simulation, the traffic analyzed is background traffic, the weight vector of TOPSIS-1 and TOPSIS-2 are displayed in figure 2 and the weight vector of each network such as WIFI, WIMAX, and UMTS which calculated by TOPSIS-3 and TOPSIS-4 are displayed in figure 3 and figure 4 respectively.

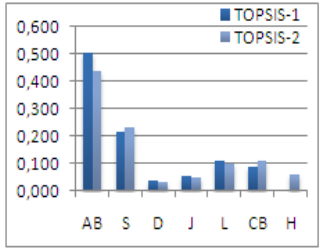
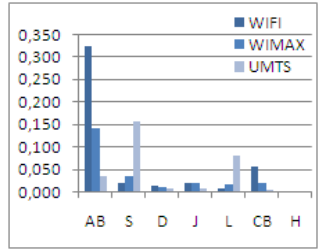
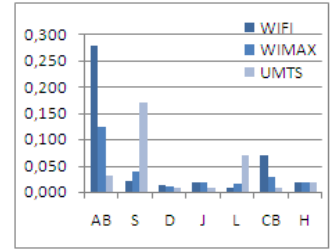

Figure 2. Weights of TOPSIS-1 and TOPSIS-2

Figure 3. Weights of TOPSIS-3

Figure 4. Weights of TOPSIS-4

### 5.1.1. Ranking abnormality

Figure 5. shows that TOPSIS-1 method reduces the risk to have this problem with a value of 33%, and TOSIS-2 method and TOPSIS-3 provide the same value for reducing the risk with a value of 25%. While TOPSIS-4 method reduces the risk with a value of 8%.

So for background traffic, TOPSIS-4 method based on differentiated weight and history attribute can reduce the ranking abnormality problem better than the all algorithms such as TOSIS-1, TOPSIS-2 and TOSIS-3, in addition the TOPSIS-2 and TOPSIS-3 which taking into consideration the differentiate weight and history attribute respectively reduce the ranking abnormality problem better than the classical network selection based on TOPSIS-1.

### 5.1.2. Number of handoffs

Figure 6. shows that TOPSIS-1 method reduces the number of handoffs with a value of 42%, and TOPSIS-2 method and TOPSIS-3 method provide the same value of the number of handoffs, the value is 25%. While TOPSIS-4 method reduces the number of handoffs with a value of 8%.

So for background traffic, TOPSIS-4 method based on differentiated weight and history attribute can reduce the number of handoffs better than the all algorithms such as TOPSIS-1, TOPSIS-2 and TOPSIS-3, in addition the TOPSIS-2 and TOPSIS-3 which taking into consideration the differentiate weight and history attribute respectively reduce the number of handoffs better than the classical network selection based on TOPSIS-1.

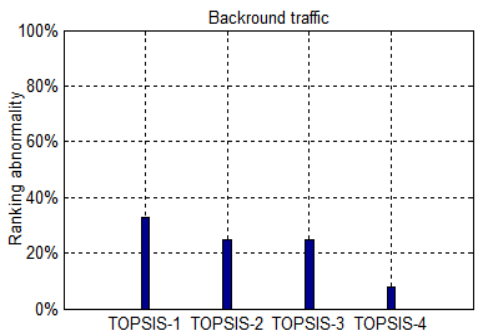 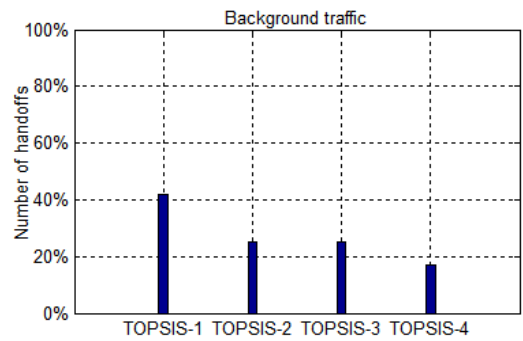

Figure 5. Average of ranking abnormality          Figure 6. Average of number of handoffs

## 5.2. Simulation 2

In this simulation, the traffic analyzed is conversational traffic, the weight vector of TOPSIS-1 and TOPSIS-2 are displayed in figure 7 and the weight vector of each network such as WIFI, WIMAX, and UMTS which calculated by TOPSIS-3 and TOPSIS-4 are displayed in figure 8 and figure 9 respectively.

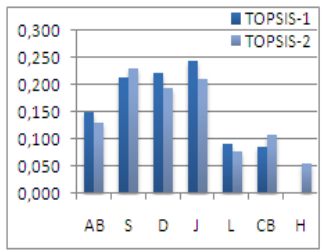 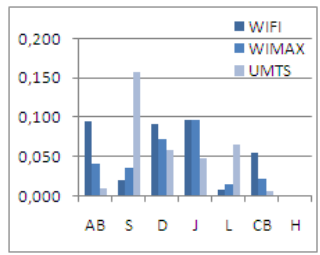 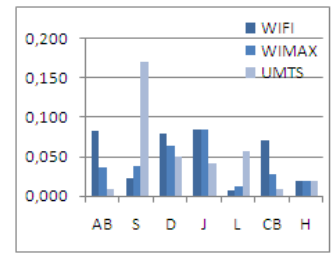

Figure 7. Weights of TOPSIS-1 and TOPSIS-2          Figure 8. Weights of TOPSIS-3          Figure 9. Weights of TOPSIS-4

### 5.2.1. Ranking abnormality

Figure 10. shows that TOPSIS-1 method reduces the risk to have this problem with a value of 25%, and TOSIS-2 method and TOPSIS-3 provide the same value for reducing the risk with a value of 17%. While TOPSIS-4 method reduces the risk with a value of 8%.

So for conversational traffic, TOPSIS-4 method based on differentiated weight and history attribute can reduce the ranking abnormality problem better than the all algorithms such as TOPSIS-1, TOPSIS-2 and TOPSIS-3, in addition the TOPSIS-2 and TOPSIS-3 which taking into consideration the differentiate weight and history attribute respectively reduce the ranking abnormality problem better than the classical network selection based on TOPSIS-1.

### 5.2.2. Number of handoffs

Figure 11. shows that TOPSIS-1 method reduces the number of handoffs with a value of 50%, and TOPSIS-2 method and TOPSIS-3 method provide the same value of the number of handoffs, the value is 42%. While TOPSIS-4 method reduces the number of handoffs with a value of 8%.

So for conversational traffic, TOPSIS-4 method based on differentiated weight and history attribute can reduce the number of handoffs better than the all algorithms such as TOPSIS-1, TOPSIS-2 and TOPSIS-3, in addition the TOPSIS-2 and TOPSIS-3 which taking into consideration the differentiate weight and history attribute respectively reduce the number of handoffs better than the classical network selection based on TOPSIS-1.

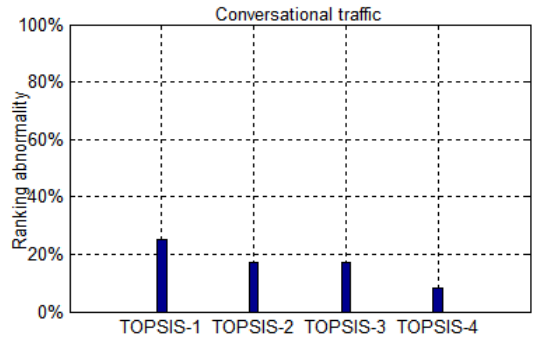
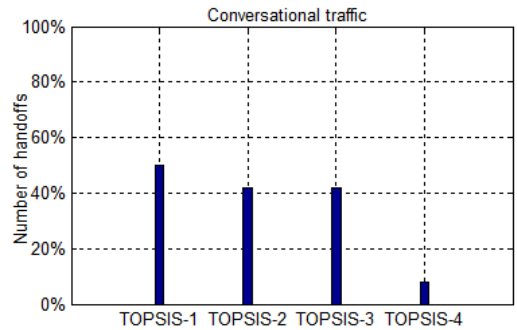

Figure 10. Average of ranking abnormality  Figure 11. Average of number of handoffs

### 5.3. Simulation 3

In this simulation, the traffic analyzed is interactive traffic, the weight vector of TOPSIS-1 and TOPSIS-2 are displayed in figure 12 and the weight vector of each network such as WIFI, WIMAX, and UMTS which calculated by TOPSIS-3 and TOPSIS-4 are displayed in figure 13 and figure 14 respectively.

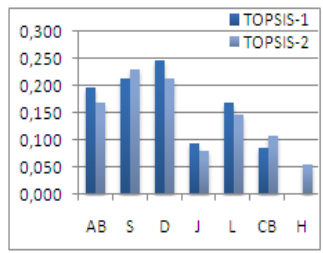
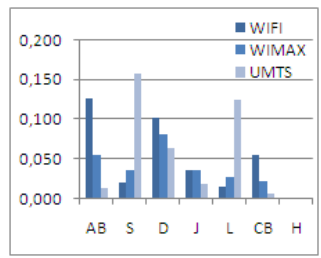
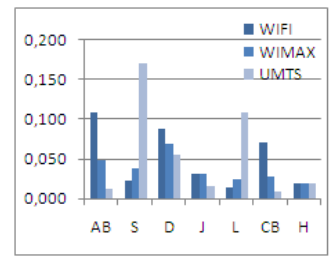

Figure 12. Weights of TOPSIS-1 and TOPSIS-2  Figure 13. Weights of TOPSIS-3  Figure 14. Weights of TOPSIS-4

### 5.3.1. Ranking abnormality

Figure 15. shows that TOPSIS-1 method reduces the risk to have this problem with a value of 25%, and TOSIS-2 method and TOPSIS-3 provide the same value for reducing the risk with a value of 17%. While TOPSIS-4 method reduces the risk with a value of 8%.

So for interactive traffic, TOPSIS-4 method based on differentiated weight and history attribute can reduce the ranking abnormality problem better than the all algorithms such as TOPSIS-1, TOPSIS-2 and TOPSIS-3, in addition the TOPSIS-2 and TOPSIS-3 which taking into consideration the differentiate weight and history attribute respectively reduce the ranking abnormality problem better than the classical network selection based on TOPSIS-1.

### 5.3.2. Number of handoffs

Figure 16. shows that TOPSIS-1 method reduces the number of handoffs with a value of 33%, and TOPSIS-2 method and TOPSIS-3 method provide the same value of the number of handoffs, the value is 25%. While TOPSIS-4 method reduces the number of handoffs with a value of 8%.

So for interactive traffic, TOPSIS-4 method based on differentiated weight and history attribute can reduce the number of handoffs better than the all algorithms such as TOPSIS-1, TOPSIS-2 and TOPSIS-3, in addition the TOPSIS-2 and TOPSIS-3 which taking into consideration the differentiate weight and history attribute respectively reduce the number of handoffs better than the classical network selection based on TOPSIS-1.

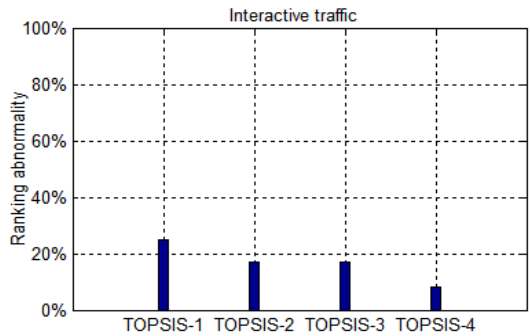
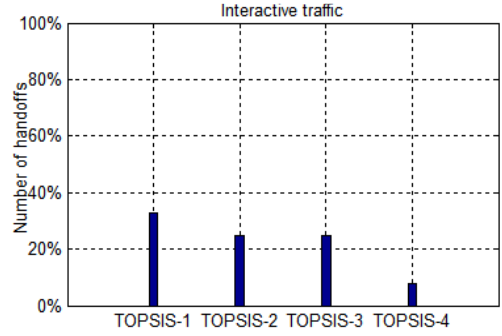

Figure 15. Average of ranking abnormality    Figure 16. Average of number of handoffs

### 5.4. Simulation 4

In this simulation, the traffic analyzed is streaming traffic, the weight vector of TOPSIS-1 and TOPSIS-2 are displayed in figure 17 and the weight vector of each network such as WIFI, WIMAX, and UMTS which calculated by TOPSIS-3 and TOPSIS-4 are displayed in figure 18 and figure 19 respectively.

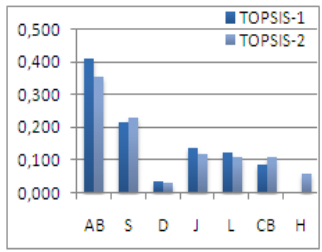
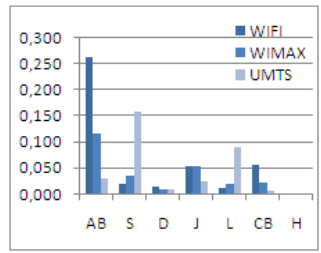
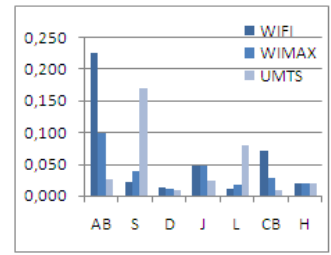

Figure 17. Weights of TOPSIS-1 and TOPSIS-2

Figure 18. Weights of TOPSIS-3

Figure 19. Weights of TOPSIS-4

### 5.4.1. Ranking abnormality

Figure 20. shows that TOPSIS-1 method reduces the risk to have this problem with a value of 42%, TOPSIS-2 method reduces the risk with a value of 33% and TOPSIS-3 method reduces the risk with a value of 25%. While TOPSIS-4 method reduces the risk with a value of 17%.

So for streaming traffic, TOPSIS-4 method based on differentiated weight and history attribute can reduce the ranking abnormality problem better than the all algorithms such as TOPSIS-1, TOPSIS-2 and TOPSIS-3, in addition the TOPSIS-2 and TOPSIS-3 which taking into consideration the differentiate weight and history attribute respectively reduce the ranking abnormality problem better than the classical network selection based on TOPSIS-1.

### 5.4.2. Number of handoffs

Figure 21. shows that TOPSIS-1 method reduces the number of handoffs with a value of 58%, TOPSIS-2 method reduces the number of handoffs with a value of 50% and TOPSIS-3 method reduces the number of handoffs with a value of 33%. While TOPSIS-4 method reduces the number of handoffs with a value of 25%.

So for streaming traffic, TOPSIS-4 method based on differentiated weight and history attribute can reduce the number of handoffs better than the all algorithms such as TOPSIS-1, TOPSIS-2 and TOSIS-3, in addition the TOPSIS-2 and TOPSIS-3 which taking into consideration the differentiate weight and history attribute respectively reduce the number of handoffs better than the classical network selection based on TOPSIS-1.

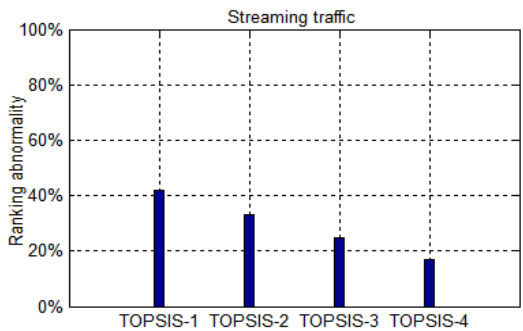
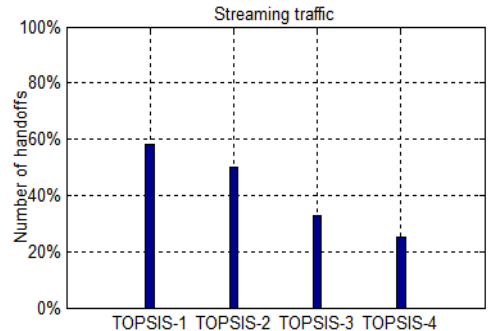

Figure 20. Average of ranking abnormality      Figure 21. Average of number of handoffs

## 6. CONCLUSIONS

In this work, we have proposed an intelligent network selection strategy namely TOPSIS-4. This strategy combines two MADM methods such as ANP method and TOPSIS method, the ANP method is applied to find the differentiate weights of available networks by considering each criterion and the TOPSIS method is used to rank the available networks. In addition the proposed strategy takes into consideration a new attribute namely history. This one helps to deal with the ping pong effect by reducing the number of handoffs.

The simulation results show that, our method based on TOPSIS-4 can reduce the ranking abnormality problem better than all algorithms such as TOPSIS-1, TOPSIS-2 and TOPSIS-3 according to all four traffic classes namely background, conversational, interactive and streaming. In the other hand for all traffic classes TOPSIS-4 method provides best performance concerning the number of handoffs than TOPSIS-1, TOPSIS-2 and TOPSIS-3.

Finally we deduce that the introducing of the differentiated weight (TOPSIS-2) or the history criterion (TOPSIS-3) in the network selection decision allows to get the best performance concerning the two aspects namely ranking abnormality and number of handoffs than the classical network selection decision (TOPSIS-1).